\documentclass[useAMS,usenatbib]{mn2e}

\title[Weyl conformastatic perihelion advance]{Weyl conformastatic perihelion advance}
\author[Abra\~{a}o J. S. Capistrano, Waldir L. Roque and Rafael S. Valada]
{Abra\~{a}o J. S. Capistrano$^{1}$\thanks{E-mail:abraao.capistrano@unila.edu.br},
Waldir L. Roque$^{2}$\thanks{E-mail:roque@ci.ufbp.br} and
Rafael S. Valada$^{3}$\thanks{E-mail:rafaelsvalada@ulbra.edu.br}\\
$^{1}$Federal University of Latin-American Integration, Technological Park of Itaipu, 85867-670, P.o.b: 2123, Foz do igua\c{c}u-PR, Brazil\\
$^{2}$Department of Scientific Computation, Federal University of Para\'{i}ba, Jo\~{a}o Pessoa, Brazil\\
$^{3}$Physics and Mathematics Unit, Lutheran University of Brazil, Porto Alegre, Brazil.}

\begin{document}

\date{in original form 2014 July 09}


\maketitle

\label{firstpage}

\begin{abstract}
In this paper, we examine a static gravitational field with axial symmetry over probe particles in the solar system. Using the Weyl conformastatic solution
as a model, we find a non-standard expression to perihelion advance due to the constraints imposed by the topology of the local gravitational field. We show that the application of the slow motion condition to the geodesic equations without altering Einstein's equations does not necessarily lead to the Newtonian limit; rather it leads to an intermediate nearly-newtonian gravitational stage, which can be applied to astrophysical problems in solar scale. We apply the model to the perihelion advance of inner planets and minor objects (NEO asteroids and the comets). As a result, we obtain an expression of a non-standard relativistic precession that reveals a close agreement to observational data calibrated with the Ephemerides of the Planets and the Moon (EPM2011). The study of perihelion advance of eight small celestial bodies (asteroids and comets) is also considered.
\end{abstract}

\begin{keywords}
nearly newtonian limit - perihelion precession - Weyl solutions.
\end{keywords}

\section{Introduction}
Usually in general relativity (GR) when slow motions are described,
the first logical approach to a classical gravitational theory is to
reduce Einstein's equations to the Newtonian limit. That limit
can be reached by many methods, such as the \emph{Parameterized
Post-Newtonian (PPN) approximation} which can be seen more as a
\emph{stage} of the newtonian approximation. This method is basically constituted of terms with superior orders in
the metric of a fallen test particle. In the PPN approximation, the metric tensor is generated
through matter distribution and hypothesized under conditions of weak gravitational field and low light speed
($v\ll c$). The arbitrary potential's coefficients are the well-known PPN parameters.

In order to avoid such parameters, in a different approach, we study the possibility of an application of GR to slow motion focusing on the geodesic equations alone. This paper aims at showing that the application of the slow motion condition to the geodesic equations without altering Einstein's equations will produce a different dynamics. As a result, it will not necessarily lead to the Newtonian limit. This solution has to do with shape, topology or aspects of symmetry of the gravitational field. In this respect, we use the Weyl metric that describes a cylindrical symmetry, motivated mainly on the fact that the solar system has an axial symmetry. From the geometrical point of view the Weyl cylindrically symmetric solution is diffeomorphic to the Schwarzschild solution and also is asymptotically flat as shown in \citep{weyl,rosen,zipoy,gau,steph}. For applications in astrophysics, the Weyl metric is fundamentally relevant as well as several non-asymptotically flat metrics, for instance, in the study of physics of black-holes, stellar evolution and galaxies as well as several studies of relativistic effects on solar system scale \citep{antonio,antonio2,katz,vogt,ujevic,ujevic2,sitarski,roberts}.

In the following section, we make a brief review of the nearly-newtonian limit. In the third section, we study the gravitational field produced by the Weyl's metric defining the coefficients of the metric. In the fourth section, the orbit equations and calculations of a non-standard expression for the perihelion advance are shown. To this end, we study the Weyl conformastatic solution (in the sense that the potential $\lambda(r,z)=0$) and obtain an expression for an orbit equation. Moreover, we study the perihelion advance for inner planets and the precession of asteroids and comets. Particularly, the relativistic effects in asteroids and comets have been explored in literature \citep{sitarski,Shahid} and advances have been made in monitoring near-Earth objets (NEOs) using radar astronomy techniques \citep{Yeomans1,Yeomans2,Margot}.

In this work, we study NEOs such as 1566 Icarus \citep{Shapiro}, 1862 Apollo, 2101 Adonis asteroids, 433 Eros and 3200 Phaethon. In addition, two Jupiter family comets (26P/Grigg-Skjellerup and 22p/Kopff) and an Encke-type comet (2p/Encke comet) are also studied. For the perihelion of inner planets we compare with the observational data \citep{nambuya,Wilhelm} calibrated with the Ephemerides of the Planets and the Moon (EPM2011) \citep{Pitjeva,Pitjev}. For the perihelion of asteroids, we compare our results with the observational data from both optical and radar observations \citep{Shahid}. For the comets, we compare the results with the numerical results based on Painlev\'{e} coordinates to one-body Schwarzschild problem \citep{sitarski}. Finally, we make the final remarks in the conclusion section.

\section{The nearly-newtonian regime}

The most classical usage of the PPN approximation was done to explain the \emph{Mercury's perihelion
advance} \citep{Nobili}. However, the PPN formalism is not the only
tool available that describes slow motions, there is another stage
that also involves space-time curvature. Infeld and Plebanski \citep{infeld} made a very interesting and
consistent demonstration of Eintein's equations through Newton's equations. They
showed how geodesic equations are contained in Einstein's
equations. Starting from Newton's equations, they took successive
approximations of the metric with the parameter $v/c \ll 1$. As a result, once the geodesic equations are built, Newton's
equations postulate of motion could be dispensed. Even though Einstein conjectured this idea in 1915,
this procedure was not clear at that time. Today it is known that a specific choice of parameters has to be made for
a specific ending. There are other parameters, not velocity related, as for example, the weak field
limit that gives the linear gravitational wave equation or the Schwarzschild weak field with the
parameter  $\frac{1}{r}$. Thus, the Newtonian equation of motion
appears in the limit of GR as an option which we need to impose the weak field condition to obtain Einstein's equations.

On the other hand, if our concern is the connection, it is well known that geodesic equations are linear in terms of the connection and
quadratic in terms of Einstein's equations. This relation can
affect the influence of the gravitational field in GR imprinting
qualitative effects on solutions.  For instance, if we only use the
geodesic equation under the hypothesis of slow motion and weak
gravitational field conditions, the field will act more smoothly,
since the connection is linear in the geodesic equations. However, if now we focus our attention on Einstein's equations,
something different will occur, since the connection is of the
fourth power on this particular set of equations. As a result, this means
that a gravitational field originated only from geodesic
equations will have different characteristics. This ``intermediate'' gravitational
regime will be located somewhere in between the gravitational filed created by Einstein's equations, which is stronger, and the newtonian field, which is weaker.
This ``new'' gravitational field is denominated \emph{nearly-Newtonian} field as stated in \citep{wheeler}.

As an example, if we consider a free-falling slow moving particle its
gravitational field will have additional increments, such as
$$ g_{\mu\nu} \approx \eta_{\mu\nu} +\delta h_{\mu\nu}+ (\delta
 h_{\mu\nu})^2 +\cdots\;.$$
Because of this additional increments, we recover the strength of the gravitational field. For this matter, the low velocity hypothesis is the only
valid one, since the field is stronger now. If this process is interrupted by an external force in a spherically symmetric matter distribution concurring with
Einstein's and geodesic equations, the newtonian gravitational field
can be restored through Poisson's equation $R^{\alpha}{}_{\alpha}=\nabla^2 \phi= 4\pi G \rho$. Conversely, if we neglect the influence of any external action, this process
goes on naturally, so we can sum all the metric's $g_{\mu\nu}$ perturbations from $\delta h_{44}=0$ to a finite value $\delta h_{\mu\nu}$. Moreover, integrating from 0 to $\delta h_{44}$, we obtain
$$\Phi_{nN}=-\frac{1}{2}\int^{\delta h_{44}}_{0}d(\delta h_{44}')=\;-\frac{1}{2}\delta h_{44}\;.$$
Hence, we can find the  \emph{nearly}-newtonian gravitational potential $\Phi_{nN}$ given by
\begin{equation}\label{eq:newpot}
\Phi_{nN}=-\frac{c^2}{2}(1+g_{44})\;,
\end{equation}
which is a similar equation for a spherically symmetric field but with a different qualitative interpretation. We stress that the $g_{44}$ metric component is an exact and  non-approximated solution.

One can insist that the same \emph{nearly}-newtonian potential found in eq.(\ref{eq:newpot}) cannot be obtained if we had just
taken the limit $v<<c$ \emph{a priori}. There is an error in this rationalization because of taking GR on its complete structure, the motion must not be given by Newton's laws, but by the geodesic equation which is a non-linear
equation of motion of the object at hand such that
$$\frac{dv^{i}}{d\tau}+\Gamma^{i}_{jk}v^{j}v^{k}=0\;.$$
At the end of the process if we apply the slow motion condition $v<<c$ to the system
altogether with the metric tensor expansion, with the parameter
$v/c$, we simply obtain Newton's theory and obtain the following
potential
$$\Phi_{nN}=-\frac{M}{r}\;,$$
which reduces to the newtonian potential that does not explain the perihelion advance. In terms of comparison,
we expect that the \emph{nearly}-newtonian potential can be used to give a description of the perihelion precession rather than the PPN approximation and extending it also to non-standard systems.

\section{Weyl's gravitational field}
Besides of its historical relevance as one of the main tests for a gravitational theory candidate, the study of the perihelion advance plays an important role on the development of gravitational physics. In recent years there has been a renewed interest in the perihelion advance and several proposals have been worked using it as one of the fundamental solar system tests, such as, e.g, the modification of Newtonian Dynamics (MOND)\citep{schimit}, azimuthally symmetric theory of gravitation based on the study of Poisson equation \citep{nambuya}, Kaluza-Klein five-dimensional gravity \citep{wesson}, Yukawa-like Modified Gravity \citep{lorenzo}, Horava-Lifshitz gravity \citep{harko}, brane-world models and variants \citep{mak,cheung,sumanta,sepangi,lorio1,lorio2} and also in the PPN framework and outside it giving rise to extra-perihelion precessions and approaches in the weak field/slow motion limits \citep{Avalos,Arakida,Adkins,Deliseo,Deng,Feldman,Kalinowski,Liang,Li,lorio3,lorio4,lorio5,lorio6,lorio7,lorio8,lorio9,lorio10,lorio11,lorio12,lorio13,Ruggiero,Xie,Wilhelm}.

On the mechanism we are going to show, we consider the effects in a single plane of
orbit. This consideration is compatible with the observed movement of the
planets around the Sun limited to the plane of orbits. Thus, we can consider the Sun in the center of the circular basis of the cylinder and a planet (or a small celestial object) as a particle with mass \emph{m} orbiting by its edge. This cylinder can be described by
Weyl's line element \citep{weyl}
\begin{equation}
ds^{2}=e^{2\left(\lambda-\sigma\right)}dr^{2}+r^{2}e^{-2\sigma}d\theta^{2}+e^{2(\lambda-\sigma)}dz^{2}-e^{2\sigma}dt^{2}\;,
\end{equation}
where $\lambda=\lambda(r,z)$ and $\sigma=\sigma(r,z)$. The exterior
gravitational field in the cylinder outskirts is given by Einstein's
vacuum equations
 \begin{eqnarray}
 && - \lambda_{,r} + r\sigma_{,r}^{2} -r
\sigma_{,z}^{2} =0\;,
\label{eq:first}\\
 &&\sigma_{,r}+r\sigma_{,rr}+r\sigma_{,zz}=0\;, \label{eq:second}\\
&& 2r\sigma_{,r}\sigma_{,z}
=\lambda_{,z}\;.\label{eq:fourth}
\end{eqnarray}
where the terms $(,r)$, $(,z)$ and $(,,r)$, $(,,z)$ denote respectively the first and the second derivatives with respect to the variables $r$ and $z$.

It is worth noting that the original paper of Weyl showed that the cylinder solution is diffeomorphic to a Schwarzschild's solution. In this process the metric does not lose its asymptotes as shown in \citep{weyl,rosen,zipoy} and is also asymptotically flat \citep{weyl,rosen,zipoy,gau,steph}. This is a fine example of the equivalence problem in GR: How do we know that two solutions of Einstein's equations, written in different coordinates, do not describe the same gravitational field? The answer is given by the application of Cartan's equivalence problem \citep{cartan} to GR. It was shown that the Riemann tensors and their covariant derivatives up to the seventh order must be equal.

It is interesting to note that regardless of velocity arguments, the diffeomorphism invariance of GR
is also broken down by the condition that the cylinder thickness $h_{0}$ is smaller than its radius
$R_{0}$, i.e., $h_{0}<< R_{0}$ being reduced to its circular basis which can be smoothly deformed by the expansion of the metric parameters. In order to analyze effects of the lack of the diffeomorphism invariance, we start with solving the non-linear system given by eqs.(\ref{eq:first}), (\ref{eq:second}) and (\ref{eq:fourth}). Differently as proposed in \citep{antonio2} and \citep{vogt} for a mass distribution with Weyl's exact solutions of Einstein equations, we study approximate solutions of this metric by expanding its coefficient functions (or potentials). To this end, we expand the coefficients $\lambda(r,z)$ and $\sigma(r,z)$ into a MacLaurin's series such that
\begin{equation}\label{eq23}
\sigma (r,z) \approx \sigma (r,0) + z{\left. {{{\partial \sigma (r,z)} \over {\partial z}}} \right|_{z = 0}} + {z^2}{\left. {{{{\partial ^2}\sigma (r,z)} \over {\partial {z^2}}}} \right|_{z = 0}} +  \cdots,
\end{equation}

\begin{equation}\label{eq24}
\lambda (r,z) \approx \lambda (r,0) + z{\left. {{{\partial \lambda (r,z)} \over {\partial z}}} \right|_{z = 0}} + {z^2}{\left. {{{{\partial ^2}\lambda (r,z)} \over {\partial {z^2}}}} \right|_{z = 0}} +  \cdots,
\end{equation}
and considering the approximation up to the second order (to keep aspects of the non-linearity of the system), we define
\begin{equation}\label{eq25}
  \sigma (r,z) = A(r) + a(r)z + c(r){z^2},
\end{equation}
where we denote $A(r)=\sigma (r,0)$, $a(r)= {\left. {{{\partial \sigma (r,z)} \over {\partial z}}} \right|_{z = 0}}$ and $c(r)=\left.{{{{\partial ^2}\sigma (r,z)} \over {\partial {z^2}}}} \right|_{z = 0}$.

In addition, we use the same procedure as in eq.(\ref{eq25}) to the coefficient $\lambda (r,z)$ and define
\begin{equation}\label{eq25a}
  \lambda (r,z) = B(r) + b(r)z + d(r){z^2}.  	
\end{equation}
where we denote $B(r)=\lambda (r,0)$, $b(r)= {\left. {{{\partial \lambda (r,z)} \over {\partial z}}} \right|_{z = 0}}$ and $d(r)=\left.{{{{\partial ^2}\lambda (r,z)} \over {\partial {z^2}}}} \right|_{z = 0}$.

The field equation in eq.(\ref{eq:fourth}) can be written as,
\begin{equation}\label{eq26}
{\sigma_{,r}} = y \Rightarrow {\sigma _{rr}} = y' \Rightarrow y + ry' + 2rc(r)=0\;,
\end{equation}
which is linear and can be solved by a factor integration. Thus, one can find
\begin{eqnarray}\label{eq30}
\sigma (r,z) =  - \int \frac{1}{r'} \left(\int 2r'' c(r'') dr''\right)dr' +\\
 \;\;\;\;\;\;\;\;\;\; A_{1}\left( z \right)\ln (r) + A_{2}\left( z \right) &&\nonumber
\end{eqnarray}

At this point, in order to make integrable eq.(\ref{eq30}), let us consider a \emph{n}-th power law solution for $c(r'') $, such as,
\begin{equation}\label{eq31}
c(r'') = {{{c_0}} \over {{{r''}^n}}},\,\,{c_0} = const\;,\;n>0\;,	
\end{equation}
which substituting into eq.(\ref{eq30}) and after some calculations, we get
\begin{equation}\label{eq36}
	 {\sigma (r,z) = k(z)\ln (r) + {{ - 2{c_0}} \over {{{(2 - n)}^2}}}{r^{2 - n}} + {A_2}\left( z \right),\,\,n \ne 2}\;.
\end{equation}

On the other hand, taking the derivative of eq.(\ref{eq36}) with respect to $z$, we obtain
\begin{equation}\label{eq37}
{\sigma _{,z}} = k(z)_z\ln (r) + A_2(z)_z\;,
\end{equation}
which from eq.(\ref{eq25}), for the same derivative we get the result
\begin{equation}\label{eq38}
{\sigma _{,z}} = a(r) + 2c(r)z = a(r) + {{2{c_0}} \over {{r^n}}}z\;.
\end{equation}

From eqs.(\ref{eq37}) and (\ref{eq38}) we see that
\begin{equation}\label{eq39}
	{A_2}\left( z \right)_{,z} = a(r) + {{2{c_0}} \over {{r^n}}}z - k(z)_z\ln (r)\;.
\end{equation}
Noting that ${A_2}\left( z \right)$ is a function of $z$ only, then
\begin{equation}\label{eq40}
   {A_2}(z) \Rightarrow
   a(r) = {a_0}\;;
   k(z) = \frac{k_0}{2}\;\;;
   n = 0.
\end{equation}
where $a_0$ and $k_0$ are constants. Therefore,
\begin{equation}\label{eq41}
{A_2}(z)_{,z} = {a_0} + 2{c_0}z \Rightarrow {A_2}(z) = {a_0}z + {c_0}{z^2}+{c_1}\;.
\end{equation}

As the field equations  eq.(\ref{eq:first}) to  eq.(\ref{eq:second}) involve only derivatives of ${\sigma (r,z)}$, we may set   ${c_1 =0}$. For the 2nd order approximation, the final form of the coefficient ${\sigma (r,z)}$ is then
\begin{equation}\label{eq42}
{\sigma (r,z) = {{{k_0}} \over 2}\ln (r) - {{{{c_0}{r^2}} \over 2}} + {a_0}z + {c_0}{z^2}}\;.
\end{equation}

As in the same fashion for the previous development to eq.(\ref{eq25}) resulting in eq.(\ref{eq42}), we apply to eq.(\ref{eq25a}). Firstly, we must define
a \emph{m}-th power law solution for the function $d(r)$ as
\begin{equation}\label{eq43}
d(r) = {{{d_0}} \over {{r^m}}},\,\,{d_0} = const\;, \;m>0\;.
\end{equation}

Considering eqs.(\ref{eq:first}) and (\ref{eq42}), and after integrating by parts in the variable \emph{r}, we obtain
\begin{eqnarray}\label{eq45}
\lambda (r,z) = \;{{{k_0}^2}\; \over 4}\;\ln (r) - \;{k_0}{c_0}{{{r^2}} \over 2}\; +\; {c_0}^2{{{r^4}} \over 4}\\
\;\;\;\;\;\;\;\;\;\;\;\;- {\left( {{a_0} + 2{c_0}z} \right)^2}{{{r^2}} \over 2} \;+ \;{B_1}(z)\;\;,&&\nonumber
\end{eqnarray}

Moreover, to obtain a closed form for  the coefficient $\lambda(r,z)$, we need to find the function $B_1(z)$. To do so, we take the derivative of eq.(\ref{eq45}) with respect to $z$ and obtain
\begin{equation}\label{eq46}
{\lambda _{,z}} = \left( { - 2{a_0}{c_0} - 4{c_0}^2z} \right){r^2} + {B_1}{(z)_{,z}}\;,
\end{equation}
and doing the same for  eq.(\ref{eq25a}), we get
\begin{equation}\label{eq46a}
{\lambda _{,z}} = b(r) + 2d(r)z = b(r) + 2{{{d_0}} \over {{r^m}}}z \;.
\end{equation}
Thus, comparing eqs.(\ref{eq46}) and (\ref{eq46a}), one can obtain
\begin{equation}\label{eq47}
{B_1}{(z)_{,z}} = b(r) + 2{{{d_0}} \over {{r^m}}}z + 2{a_0}{c_0}{r^2} + 4{c_0}^2z{r^2}\;.
\end{equation}
Integrating eq.(\ref{eq47}) with respect to $z$, gives
\begin{equation}\label{eq49}
{B_1}(z) = \left( {{{{d_0}} \over {{r^m}}} + 2{c_0}^2{r^2}} \right){z^2} + \left( {b(r) + 2{a_0}{c_0}{r^2}} \right)z + {b_1}\;.
\end{equation}

As ${B_1}(z)$ is a function of  $z$ only, we need that
\begin{eqnarray}\label{eq50}
C_1 = \frac{d_0}{r^m} + 2{c_0}^2{r^2} = d(r) + 2{c_0}^2{r^2}\\
C_2 = b(r) + 2{a_0}{c_0}{r^2}\;,&&\nonumber
\end{eqnarray}
where ${\rm C_1}$ e ${\rm C_2}$, are constants. Thus, ${\lambda (r,z)}$ is expressed by

\begin{eqnarray}\label{eq51}
\lambda (r,z) = {{{k_0}^2} \over 4}\ln (r) - {k_0}{c_0}{{{r^2}} \over 2} - {\left( {{a_0} + 2{c_0}z} \right)^2}{{{r^2}} \over 2}\\
\;\;\;\;\;\;\;\;\;\;\;\;+ \frac{1}{4}c_0^2 r^4 + {\rm C_2}z +{\rm C_1}{z^2}\;\;,&&\nonumber
\end{eqnarray}
where it has been assumed that the constant $b_1=0$. Substituting eqs.(\ref{eq51}) and (\ref{eq42}) into the field equations eqs.(\ref{eq:first}), (\ref{eq:second}) and (\ref{eq:fourth}), we find that
\begin{equation}\label{eq52}
{\rm C_1} = {k_0}{c_0}\;,
\end{equation}
and
\begin{equation}\label{eq53}
{\rm C_2} = {k_0}{a_0}\;.
\end{equation}

Finally, the 2nd order approximation for the coefficient $\lambda(r,z)$ can be written as
\begin{eqnarray}\label{eq54}
\lambda (r,z) = {{{k_0}^2} \over 4}\ln (r) - {k_0}{c_0}{{{r^2}} \over 2} + \frac{1}{4}c_0^2 r^4 \\
\;\;\;\;\;\;\;\;\;\;\;\;- {\left( {{a_0} + 2{c_0}z} \right)^2}{{{r^2}} \over 2} + {k_0}{a_0}z + {k_0}{c_0}{z^2}.\;\;&&\nonumber
\end{eqnarray}

To complete the calculations, we need to know the relations between the parameters of the coefficients of expansion $a(r)$, $b(r)$, $c(r)$ and $d(r)$. It can be easily checked by using eqs.(\ref{eq50}), (\ref{eq52}) and (\ref{eq53}), and we have the results
\begin{equation}\label{eq55}
b(r) = {k_0}{a_0} - 2{a_0}{c_0}{r^2},
\end{equation}
and also
\begin{equation}\label{eq56}
d(r) = k_0 c_0 - 2{c_0}^2{r^{2}},
\end{equation}
which we conclude that
\begin{equation}\label{eq57}
  b(r) = \frac{a_0}{c_0} d(r)\;.
\end{equation}
It is worth nothing that for superior orders, the terms in the coefficients $\sigma(r,z)$ and $\lambda(r,z)$ turn to be redundant and can be reduced to the second order.

In order to test this assumption, we study the gravitational field produced by geodesic equation alone which can induce to an intermediate gravitational regime between Einstein's strong field and the newtonian field, known as nearly-newtonian regime \citep{wheeler,infeld}.

Since the relativistic effects generated by solar gravity is $10^{-8}$ weaker that newtonian ones \citep{yamada}, we assume that the second order of expansion of the coefficients must represent a small perturbation of the first order in such a way that ${c_0} \ll 1$ and also the usage of Weyl conformastatic solution (in the sense that the potential $\lambda(r,z)=0$) to guarantee that the resulting gravitational field produced can be enough strong to give a proper correction for the perihelion advance.

\section{Orbit equations and the perihelion advance}

Besides of calculating the Einstein's equations for Weyl metric, we must obtain an orbit equation in order to deal with the perihelion advance. To this end, we calculate the geodesics from Weyl metric and find the following components
\begin{eqnarray}\label{eq58}
  {{{{d^2}r} \over {d{s^2}}} + \left( {{\sigma_{,r}} - {\lambda_{,r}}} \right){{\left( {{{dz} \over {ds}}} \right)}^2} + \left( {2{\lambda_{,z}} - 2{\sigma_{,z}}} \right){{dr} \over {ds}}{{dz} \over {ds}}} +    \\ \nonumber
  {{e^{ - 2\lambda }}\left( {{r^2}{\sigma_{,r}} - r} \right)}{{{\left( {{{d\theta } \over {ds}}} \right)}^2} + {e^{4\sigma  - 2\lambda }}{\sigma_{,r}}{{\left( {{{dt} \over {ds}}} \right)}^2}}  & &  \\ \nonumber
  - \left( {{\sigma_{,r}} - {\lambda_{,r}}} \right){{\left( {{{dr} \over {ds}}} \right)}^2} = 0\;, & &
\end{eqnarray}
and also the following set of equations
\vspace{0.5cm}
\begin{equation}\label{eq59}
{2r{\sigma_{,z}}{{d\theta } \over {ds}}{{dz} \over {ds}} - r{{{d^2}\theta } \over {d{s^2}}} + 2r{\sigma_{,r}}{{dr} \over {ds}}{{d\theta } \over {ds}} - 2{{dr} \over {ds}}{{d\theta } \over {ds}} = 0}\;,
\end{equation}
\vspace{0.5cm}

\begin{eqnarray}\label{eq60}
  {{{{d^2}z} \over {d{s^2}}} + \left( {{\lambda_{,z}} - {\sigma_{,z}}} \right){{\left( {{{dz} \over {ds}}} \right)}^2} - \left( {2{\sigma_{,r}} - 2{\lambda_{,r}}} \right){{dr} \over {ds}}{{dz} \over {ds}} +  }& &  \\ \nonumber
  {{e^{ - 2\lambda }}{r^2}{\sigma_{,z}}{{\left( {{{d\theta } \over {ds}}} \right)}^2}+ {e^{4\sigma  - 2\lambda }}{\sigma_{,z}}{{\left( {{{dt} \over {ds}}} \right)}^2}} +  & & \\ \nonumber
   - \left( {{\sigma_{,z}}-{\lambda_{,z}}} \right){{\left( {{{dr} \over {ds}}} \right)}^2} = 0 &&
\end{eqnarray}

\begin{equation}\label{eq61}
{2{\sigma_{,z}}{{dt} \over {ds}}{{dz} \over {ds}} + {{{d^2}t} \over {d{s^2}}} + 2{\sigma_{,r}}{{dr} \over {ds}}{{dt} \over {ds}} = 0}\;.
\end{equation}

As a result, one can obtain the orbit equation
\begin{equation}\label{eq71}
{\left( {{{dr} \over {d\theta }}} \right)^2} = {e^{ - 2\lambda }}\left[ {{r^4}{e^{ - 2\sigma }}\left( {\alpha  + \beta {e^{ - 2\sigma }}} \right) - {r^2}} \right]\;,
\end{equation}
where $\alpha_0  = {1 \over {{k_2}^2}}$ and $\beta_0  = {{{k_1}^2} \over {{k_2}^2}}$ are integration constants.

Due to the structure of Weyl's metric, we only need the coefficient $\sigma$ to produce a nearly newtonian gravitational regime from the component $g_{44}$. Thus, we can consider only study a conformastatic solution \citep{antonio} for eq.(\ref{eq71}) and obtain
\begin{equation}\label{orbit}
\left(\frac{dr}{d\theta}\right)^2 =\; \left[ r^4 e^{-2\sigma} \left(\alpha_0 + \beta_0 e^{-2\sigma} \right)- r^2 \right]\;,
\end{equation}
which can be transformed into the following conformastatic orbit equation
\begin{equation}\label{orbit1}
\left(\frac{du}{d\theta}\right)^2 + u^2 =\; e^{-2\sigma} \left(\alpha_0 + \beta_0 e^{-2\sigma} \right)\;,
\end{equation}
where the new variable $u$ stands for $u= \frac{1}{r}$ and $\sigma= \sigma(u)$.

\subsection{First order approximation}
In order to make clear the physical differences between the first and second order of the coefficient $\sigma$, we present orbit equations produced by each order. Using eq.(\ref{eq42}), we stress that in the first approximation we have $c_0 =0$ and the coefficient $\sigma(r,z)|_{z=0}$ is simply reduced to
$$\sigma(r,z)|_{z=0} = \frac{k_0}{2} \ln r\;.$$

In order to obtain the resulting perihelion advance, first of all, in the same way as in \cite{harko}, we evaluate the first approximation of Weyl's conformastatic metric and find
\begin{equation}\label{orbit2}
\left(\frac{du}{d\theta}\right)^2 + u^2 = G(u)\;,
\end{equation}
where $G(u)$ is given by $\left(\alpha_0 u^{k_0} + \beta_0 u^{2k_0} \right)$ with the constants $\alpha_0$ and $\beta_0$. Deriving eq.(\ref{orbit2}), we can write
\begin{equation}\label{orbit3}
u'' + u = F(u)\;,
\end{equation}
where $F(u)$ is defined
$$F(u)= \frac{1}{2} \frac{dG(u)}{du}= \;\frac{1}{2} \left[\alpha_0 k_0 u^{k_0-1} + 2k_0 \beta_0 u^{2k_0-1} \right]\;.$$

The perihelion advance is given by
\begin{equation}\label{perih}
\delta\phi = \pi \frac{dF(u)}{du}\rfloor_{u=u_0}\;.
\end{equation}
A nearly circular orbit is given by the roots of the equation $F(u_0)=u_0$. In order to get the correct decaying law for the gravitational field (in solar system no stronger than $\sim 1/r^3$), we set $k_0=2$ and find
$$u_0= \sqrt{\frac{1-\alpha_0}{2\beta_0}}\;.$$
Thus, after using the proper equivalence with the standard gravity and neglecting the redundant terms, one can find the following expression
\begin{equation}\label{perifirstorder}
\delta\phi \approx \frac{4\pi GM}{c^2 \gamma(1-\epsilon^2)}\;,
\end{equation}
where $\gamma$ is the semi-major axis and $\epsilon$ is the eccentricity. Clearly, the expression in eq.(\ref{perifirstorder}) generates a angular deviation smaller than observed. Interestingly, eq.(\ref{perifirstorder}) resembles the gravitational lens equation.

\subsection{Second order approximation}
\begin{table*}
 \centering
 \begin{minipage}{140mm}
  \caption{Relevant proper elements for computing the perihelion precession of minor objets. Except for 1566 Icarus, the following data below can be found in JPL small body database (http://ssd.jpl.nasa.gov/sbdb.cgi). The semi-major is shown in astronomical units ($1 AU =  1.49598 \times10^{11} m$). The orbital periods are in units of years.}
  \begin{tabular}{@{}llrrrrlrlr@{}}
  \hline
   Object                                                                            &   Semi-major     &  Mean eccentricity &  Period (yr)   \\
  \hline
  1566 Icarus    \footnote{Observational data extracted from \citep{Wilhelm}.}       &   1.07792        & 0.82695       &  1.11995    \\
  1862 Apollo                                                                        &   1.47011        & 0.55993       &  1.78    \\
  2101 Adonis                                                                        &   1.87447        & 0.76381       &  2.57    \\
  433 Eros                                                                           &   1.45797        & 0.22263       &  1.76    \\
  3200 Phaethon                                                                      &   1.27112        & 0.88985       &  1.43     \\
  26P/Grigg-Skjellerup                                                               &   3.01721        & 0.64016       &  5.24    \\
  22p/Kopff                                                                          &   3.44968        & 0.54734       &  6.41    \\
  2p/Encke                                                                           &   2.21473        & 0.84823       &  3.3     \\
  \hline
\end{tabular}
\end{minipage}
\end{table*}

\begin{table*}
  \centering
  \begin{minipage}{140mm}
  \caption{Comparison between the values for secular precession of inner planets in units of arcsec/century of the standard (Einstein) perihelion precession $\delta \phi_{sch}$ \citep{Wilhelm} and the conformastatic solution $\delta \phi_{model}$. The $\delta \phi_{obs}$ stands for the secular observed perihelion precession in units of arcsec/century adapted from \citep{nambuya} by adding a supplementary precession corrections from EPM2011 \citep{Pitjeva,Pitjev}.}
  \begin{tabular}{@{}llrrrrrrr@{}}
  \hline
   Object             &    $\delta \phi_{obs}$ $(''.cy^{-1})$  &  $\delta \phi_{sch}$$(''.cy^{-1})$      &
   \multicolumn{4}{c}{$\delta \phi_{model}$$(''.cy^{-1})$ } $\beta_0$ & \\
                                                                             &&   &$\;\;\;\delta \phi^{+} $& $\;\;\;\delta \phi^{-} $&  &   \\

  \hline
  Mercury            & 43.098 $\pm$ 0.503        &   42.97817          &  43.1047    &  42.8569   & &   $1.7497\times 10^{-3}$ \\
                     &                           &                  &&  \\

  Venus              & 8.026  $\pm$ 5.016        &   8.62409         &  8.62462      & 8.62462    & & $2.10433\times 10^{-9}$ \\
                     &                           &                  & &  \\

  Earth              & 5.00019  $\pm$ 1.00038    &   3.83848          & 3.83873      & 3.83873    & & $7.79739\times10^{-8}$ \\
                     &                           &                  & &  \\

  Mars               & 1.36238  $\pm$ 0.000537   &   1.35086          & 1.35086      & 1.35086    & & $7.58002\times10^{-5}$\\
                     &                           &                  & &  \\

  \hline
 \end{tabular}
\end{minipage}
\end{table*}

For the second order we have that the constant $c_0$ plays the role of the correction term. Essentially, we use the same procedure as describe in the first order approximation. Using eq.(\ref{eq42}) calculated in $z=0$, we obtain the general expression for $F(u)$ and find
\begin{eqnarray}\label{fu}
F(u)= \frac{1}{2}\left[\alpha_0 k_0 u^{k_0-1} + 2\beta_0 k_0 u^{2k_0-1} \right] \\
\;\;\;\;\;\;\;\;\;\;+ \frac{1}{2}\alpha_0 c_0 (k_0-2)u^{k_0-3}+ 4(k_0-1)\beta_0c_0 u^{2k_0-3}\; . \nonumber
\end{eqnarray}

Interestingly, we find that the values for the $k_0$ parameter can be very constrained based on the fact that it must provide a correct decaying law in the solar system scale which is ruled essentially by newtonian potentials ($\sim\frac{1}{r}$) and smooth deviations in decaying $(\sim\frac{1}{r^2},\sim\frac{1}{r^3})$ and growing not large as $\sim r^2$. The increasing of gravitational potentials of order of $\sim r^3$, or quadratically and on, should be quite worrisome in the solar system scale.

For $k_0=0$, we can reproduce the same orbit equation as shown in \cite{roberts} in the study of the perihelion of Pluto. From $k_0\leq-1$ in orbit equation, we find a very fast growing terms that reproduces an incorrect secular shift for the perihelion. For the case when $k_0 > 2$ it produces orbit equations with high-decaying terms of order of $\mathcal{O}(u^4)$. On the other hand, for the case $k_0=2$ we find that it reproduces the correct decaying law and one can obtain the following terms
\begin{equation}\label{orbit2order}
F(u)= \frac{1}{2} \left[ (2\alpha_0 + 4\beta_0c_o)u + 4\beta_0u^3\right]\;.
\end{equation}
For $F(u_0)=u_0$, one can find the roots
\begin{equation}\label{root2}
u_0^2= \frac{1-(\alpha_0+2\beta_0c_0)}{4\beta_0}\;,
\end{equation}
and using (\ref{perih}), (\ref{orbit2order}) and (\ref{root2}), one can obtain the shifted angle $\delta\phi$ as
\begin{equation}\label{perih2}
    \delta\phi = \frac{(3-\alpha_0)\pi}{2} + 4\beta_0c_0\pi\;.
\end{equation}

In order to relate to the standard gravity, we set $(3-\alpha_0)=\frac{12 GM}{c^2 \gamma(1-\epsilon^2)}$ and find the correction for the perihelion advance
\begin{equation}\label{perih3}
    \delta\phi = \delta \phi_{sch} \pm 4\pi \beta_0 c_0\;.
\end{equation}
where we denote $\delta \phi_{sch}=\frac{6\pi GM}{c^2 \gamma(1-\epsilon^2)}$, with $\gamma$ being the semi-major axis and $\epsilon$ denotes the eccentricity of the orbits. In addition, based on the fact that $c_0 << 1$, it can be constrained setting $c_0 = \pm \frac{1}{4\pi}\;\nu$ where $\nu$ is the keplerian mean motion given by $\sqrt{\frac{GM}{\gamma^3}}$. It worth noting that two solutions are possible with the sign $\pm$ related to the disk expansion (the variation of the orbit with reference to the plane). If we conveniently set the parameter $\beta_0 << 1$, the eq.(\ref{perih3}) reproduces the same classical GR results. On the other hand, we are interested in the case that the parameter $\beta_0$ can induce a correction to the standard $\delta \phi_{sch}$ generated by PPN-parameters, so the eq.(\ref{perih3}) in principle can be applied to model non-trivial systems with non-standard perihelion deviations with only one parameter. To this end, the values of $\beta_0$ must carry some information of each studied problem and being constrained by the assumption
\begin{equation}\label{perih4}
    \beta_0 =  \epsilon^4 \sqrt{1-\epsilon^2}\;.
\end{equation}
with $\beta_0$ restricted to the interval $[0,1]$.

Thus, one can rewrite the eq.(\ref{perih3}) as
\begin{equation}\label{perih5}
    \delta\phi = \delta \phi_{sch} \pm \epsilon^4\sqrt{\frac{GM(1-\epsilon^2)}{\gamma^3}}\;.
\end{equation}
If the orbit is nearly circular, the previous equation is reduced to the standard Schwarzschild one.

Next, we apply the model to the inner planets, once we have a more reliable data in this range. The relevant proper elements (semi-major, mean eccentricity and period) for computing the perihelion precession of inner planets and 1566 Icarus asteroid were extracted from \citep{Wilhelm}. The values of constant considered in the calculations are the following: the Newtonian constant of gravitation $G= 6.67384\times10^{-11} m^3 kg^{-1}.s^{-2}$ \citep{Wilhelm}, one year $1 yr = 365.256 d$, the speed of light $c= 299792458 m/s$ \citep{Wilhelm,Bureau} and the mass of sun $M_{\odot}= 1.98853\times 10^{30} kg$.

In Table (1), we present the relevant quantities to calculation of the angular deviation for selected asteroids and comets. Due to the fact of its importance of study, we have selected NEO's of Apollo group which consists asteroid orbits that cross the Earth's orbit close to that of 1862 Apollo (with semi-major axis of order of $a > 1.0$ AU and perihelion distance $q < 1.017$ AU) being also considered potentially hazardous asteroids (PHA). For the second-largest NEOs 433 Eros is located in Amor group (it crosses the Earth's orbit close to that of 1221 Amor with $1.017 < q < 1.3$ AU). For the comets, we have selected a NEO 2p/Encke comet and two Jupiter-family comets 26P/Grigg-Skjellerup and 22p/Kopff which properties are being currently investigated \citep{Busemann,Levison,Moreno}.

As a result, we obtain  Table (2) for the values of $\beta_0$ (using eq.(\ref{perih4})) and a comparison of the different angular deviation of both observational and PPN results.

\begin{table*}
  \centering
  \begin{minipage}{140mm}
  \caption{Comparison between the values for secular precession in units of arcsec/century of the standard (Einstein) perihelion precession and the conformastatic solution $\delta \phi_{model}$ for selected NEO asteroids and comets. The $\delta \phi_{obs}$ stands for the secular observed perihelion precession in units of arcsec/century. For the comets $\delta \phi_{obs}$ is the numerical results using Painlev\'{e} coordinates to one-body Schwarzschild problem \citep{sitarski}.}
  \begin{tabular}{@{}llrrrrrr@{}}
  \hline
   Object             &    $\delta \phi_{obs}$ $(''.cy^{-1})$  &  $\delta \phi_{sch}$$(''.cy^{-1})$      &
   \multicolumn{4}{c}{$\delta \phi_{model}$$(''.cy^{-1})$ } $\hspace{-1cm}\beta_0$ & \\
                                                                             &&   &$\delta \phi^{+} $& $\delta \phi^{-} $&  &   \\

  \hline

  1566 Icarus           & 10.007  &   10.06128 \footnote{Extracted from \citep{Wilhelm}.}    &  10.9240   &  9.19979    & &   0.262946\\
                        &                       &                  && \\

  1862 Apollo           & 2.1239  &   2.133                                                  &  2.23863   & 2.02809     & &   0.081442\\
                        &                       &                  & &   \\

  2101 Adonis           & 1.9079  &   1.912                                                  &  2.05192   &  1.77796   & &   0.219684   \\
                        &                       &                  & &   \\

  433 Eros              & 1.60    &   1.57317                                                &  1.57634  & 1.56999        & &   0.002395 \\
                        &                       &                  & & \\

  3200 Phaethon         & 10.1    &   10.1201                                                 &  10.6921  & 9.54815         & &   0.286071 \\
                        &                       &                  & & \\

  26P/Grigg-Skjellerup  & 0.54    &   0.4106                                               &  0.430490  & 0.391913    & &   0.129018  \\
                        &                       &                  & & \\

  22p/Kopff             & 0.288   &   0.2474                                               &  0.255405  & 0.240378    &&    0.075112   \\
                        &                       &                  &  &              \\

  2p/Encke              & 1.9079               &   1.868                                     &  1.97788  & 1.77058        & &   0.274174 \\
                        &                       &                  & & \\

  \hline
 \end{tabular}
\end{minipage}
\end{table*}

As Tables (2) and (3) indicate, the calculated angular deviation $\delta \phi$ from conformastatic solution is consistent with observations. For inner planets, it is clear that the planets with low eccentricity generates a low value for $\beta_0$ and essentially reproduce the same results as PPN approximation does, except for Mercury which presents a larger eccentricity among the planets and the correction for perihelion advance appears in good agrement with observations and other works in literature \citet{Anderson,Shapiro2,nambuya,harko}. Moreover, we find a very small $\beta_0$, i.e, $\beta_0 << 1$ for the cases of Venus, Earth and Mars, and we do not observe any difference in $\delta \phi^{+} $ and $\delta \phi^{-} $ solutions. In this scale, the values of $\beta_0$ seem to suggest that the larger is $\beta_0$, the larger is the perihelion advance. This similar pattern also appears in the asteroid/comet scale. We point out that $\beta_0$ essentially depends on the eccentricity of the orbit.

On the other hand, in the asteroids and comets scale, we find a tighter constraint for $\beta_0$ in the range $0< \beta_0 < 0.3$ providing solutions compatible with the observed perihelion for the selected asteroids and in the case of comets the solutions are also compatible with the numerical results using Painlev\'{e} coordinates \citep{sitarski}. It is interesting to point out that in the case of 1566 Icarus and 3200 Phaethon have a curious fact, they have similar eccentricities and perihelion advances. We  conjecture if orbiting bodies with roughly same dynamical and cinematical characteristics with similar values of $\beta_0$ could have roughly the same perihelion advance. Nevertheless, doing such affirmation is still premature and must be investigated further.

It is worth noting that the $\delta \phi_{sch}$ deviation from the standard Einstein precession can be improved mainly on the small celestial bodies where the relativistic effects can also be observed and their eccentricity are larger. Conversely, we show that a simply analysis of the non-linearities and qualitative effects of the gravitational field, we obtain non-standard results without proposing a different theory of gravitation just taking into account the shape, the topology and the symmetry aspects of the gravitational field.

\section{Final remarks}
In the \emph{nearly}-newtonian potential, which is originated from
the impositions made on geodesic equations, the $g_{44}$ metric
component carries the non-linearity of Einstein's equations. This approximation is essentially an application of GR to slow motion. Note that the
equivalence principle remains valid but the same does not occur with the
generalized covariance which is broken when the condition $v\ll c$
is postulated. This means that making the choice of an adequate
geometry becomes a very important matter, since the diffeomorphism
transformations are not valid anymore. In this respect, it should be noted that the Weyl cylindrical
solution can be transformed to the Schwarzschild's solution by a diffeomorphism \citep{rosen}. However, we cannot apply
such transformation here because the diffeomorphism invariance has been lost.

Indeed, the \emph{nearly}-newtonian limit is quite different $\;$from$\;$ the
$\;$$\;$PPN approximation in which a choice of parameters is
necessary to define the arbitrary potential's coefficients. Besides of
losing general covariance as a consequence of the slow motion
condition in the \emph{nearly}-newtonian domain, only one component
of the metric has a direct contribution to the motion.

We know from the study of dynamical systems that the
non-linearities imprint qualitative effects on the orbits of their
solutions which was shown by the conformastatic coordinates used here. The second order of the expansion of the coefficients $\sigma$ and $\lambda$ was enough to obtain the appropriated gravitational field, once the superior orders in the course of this study revealed that they can be reduced to the second order considering the disk expansion in mind. In the perihelion case, we obtained a non-standard expression for the perihelion precession that can be extended to the analysis of extra-solar systems. Rather, the method itself can be applied to any problem that the strength of the gravitational field is constrained by the topology of the problem.

In order to test the model we use the observational data adapted from \citep{nambuya} with supplementary precession corrections from EPM2011 \citep{Pitjeva,Pitjev} to refine our results for the perihelion advance of inner planets. For asteroids and comets also have shown a good agreement with the database \citep{Wilhelm}\citep{sitarski} used for comparisons. In all cases, we have found that the value of $\beta_0$ is related to the perihelion advance in directly proportional form.

The main advantage of this process resides in its simplicity. In addition, the topological nature of the problem is now an important character which is not take into account in the PPN approximation. As a result, we have obtained a model of only one parameter which can be easily constrained. Some non-standard examples were studied regarding small celestial body objects. These objects in general are very difficult to model and we have obtained a good agreement to observations improving the results of the PPN approximation. All the results  we have presented are essentially in the realm of GR. In this context, additional assumptions or modifications of the standard gravity are not needed. As future perspectives, an extended analysis of the deviation of light, radar echo and gravitational lens in spheroidal metrics are currently in progress.

\label{lastpage}

\end{document}